\documentclass[aps,prd,superscriptaddress,amsfonts,amssymb,amsmath,eqsecnum,nofootinbib,twocolumn,floatfix]{revtex4}

 \usepackage{graphicx}
 \usepackage{dcolumn}
 \usepackage{bm}
 \usepackage[T2A]{fontenc}
 \usepackage{mathptmx}
 \usepackage{mathtools}
 \usepackage{amsmath}
 \usepackage{amssymb}
 \usepackage{ulem}
 \usepackage{caption}
 \usepackage{subcaption}
 \usepackage{indentfirst}
 \usepackage{siunitx}
 \usepackage{chemformula}
 \DeclareSIUnit\gauss{G}

 \usepackage{color,graphicx}
 \usepackage[utf8]{inputenc}
 \usepackage[T1]{fontenc}
 \usepackage[russian,english]{babel}

 \definecolor{darkblue}{rgb}{0,0,0.7}
\definecolor{darkred}{rgb}{0.7,0,0}
\definecolor{darkgreen}{rgb}{0,0.4,0}
\usepackage[unicode, colorlinks, citecolor=darkblue, linkcolor=darkred, urlcolor=blue]{hyperref}

\allowdisplaybreaks
\begin{document}

\date{\today}
	
\title{Magneto-optical effects in a high-$Q$ whispering-gallery-mode resonator with a large Verdet constant}

\author{Andrey Danilin\footnote{Corresponding author: a.danilin@rqc.ru}}
\affiliation{Faculty of Physics, M.V. Lomonosov Moscow State University, Leninskie Gory, Moscow 119991,  Russia}
\affiliation{Russian Quantum Center, 143026 Skolkovo, Russia}

\author{Grigorii Slinkov}
\affiliation{Faculty of Physics, M.V. Lomonosov Moscow State University, Leninskie Gory, Moscow 119991,  Russia}

\author{Valery Lobanov}
\affiliation{Russian Quantum Center, 143026 Skolkovo, Russia}

\author{Kirill Min'kov}
\affiliation{Russian Quantum Center, 143026 Skolkovo, Russia}

\author{Igor Bilenko}
\affiliation{Faculty of Physics, M.V. Lomonosov Moscow State University, Leninskie Gory, Moscow 119991,  Russia}
\affiliation{Russian Quantum Center, 143026 Skolkovo, Russia}

\begin{abstract}
 We have studied magneto-optical effects in an optical whispering-gallery-mode resonator (WGMR) manufactured from a Faraday-rotator material with, to the best of our knowledge, the record quality factor ($Q = 1.45\times10^8$) achieved for such materials. We have experimentally measured the eigenfrequencies’ deviation amplitude under the application of an external magnetic field and demonstrated the polarization plane declination over the light path. An analytical model for arbitrary magnetic field geometries in magneto-optic birefringent WRMRs has been developed.
 \vspace{0.5 cm}
 
 © 2021 Optical Society of America. One print or electronic copy may be made for personal use only. Systematic reproduction and distribution, duplication of any material in this paper for a fee or for commercial purposes, or modifications of the content of this paper are prohibited. Hyperlink to the online abstract in the OSA Journal: \href{https://doi.org/10.1364/OL.422322}{doi.org/10.1364/OL.422322}.
\end{abstract}

\maketitle

\section{Introduction}
Unprecedented properties of whispering-gallery-mode resonators (WGMRs) \cite{BGII, Ultra_high_Q_crystalline_microcavities, WGM_microresonators_sensing_lasing, Nonlinear_and_quantum_optics, Optical_resonators_part1, Optical_resonators_part2}, including extraordinary quality factors and extremely small mode volume defined their use in numerous contemporary applications, such as photonics, nonlinear optics, and precise measurements \cite{Kilohertz_optical_resonances}. WGMRs are widely used to realize various effective detectors \cite{Acoustic_sensor, Temp_sensor, Humidity_sensor, fiberstrain}. Resonators made of electro-optical materials (e.g. $LiNbO_3$, $LiTaO_3$) allow precise resonant-frequency control by means of electric field application and are used as high-speed modulators \cite{electroopticalmodulation}.  In contrast, WGMRs from magnetic field-sensitive materials are much less studied, though outstanding original sensors based on magnetostriction were demonstrated \cite{yu2016optomechanical}. 

In this work, we aim to investigate the possibility of magneto-optical effects utilization for the purpose of WGMR-enhanced magnetic field sensing. The Terbium Gallium Garnet ($Tb_3 Ga_5 O_{12}$, TGG) has been chosen the WGMR material  for two reasons. First, it shows a strong magneto-optical effect (Faraday polarization rotation), bearing an outstanding Verdet constant $\nu$ of $\SI[per-mode=symbol]{-111}{\radian\per\tesla\per\metre}$ at $\SI{670}{\nano\metre}$ wavelength \cite{TGG_pulsed, TGG_temp_wave}. For example, pure silica has $\nu=\SI[per-mode=symbol]{2.80}{\radian\per\tesla\per\metre}$ at the same wavelength \cite{SilicaFibreFaraday}. Second, it has low intrinsic losses ($\alpha <\SI{0.001}{\per\centi\metre}$) at the same wavelength, which is promising for a high $Q$-factor of a WGMR. This unique combination of useful properties defined its application in optical isolators. TGG is reported to have the figure of merit $\nu/\alpha\sim\SI[per-mode=symbol]{1100}{\radian\per\tesla}$ at $\SI{670}{\nano\metre}$ \cite{TGG_TSLAG_FOM}. Its close competitor, yttrium iron garnet (YIG), shows $\nu/\alpha=\SI[per-mode=symbol]{2.51}{\radian\per\tesla}$ at $\SI{780}{\nano\metre}$ \cite{YIG_FOM} due to the higher intrinsic losses, which limits WGMR performance. 

In this study, we have predicted and observed experimentally the presence of the Faraday effect-induced eigenfrequency deviation -- the WGM resonant wavelength dependence on the external magnetic field magnitude. Also, we have implemented a two-port polarization control setup and confirmed that the polarization plane experiences path-dependent declination, which nullifies for a complete roundtrip.
\begin{figure}[bht]
        \centering
        \includegraphics[width=0.99\linewidth]{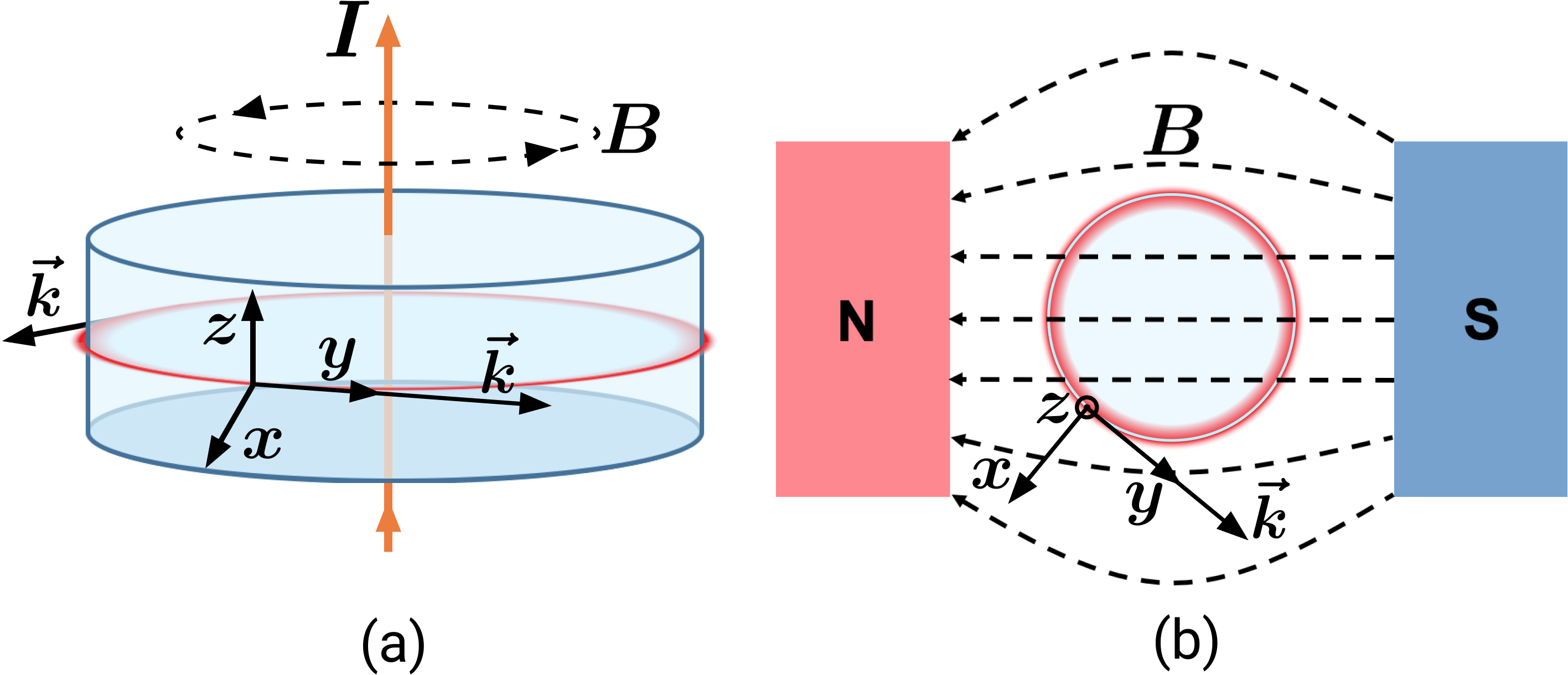}
        \caption{Light-magnetic field interaction geometries in an axisymmetric WGMR (optical path shown in red): (a) collinear configuration -- magnetic field $\boldsymbol{B}$ produced by current $\boldsymbol{I}$ is co-aligned with the wave vector along the light's path; (b) transverse configuration -- the magnetic field is uniform along the light's path.}
        \label{fig:field_orientation}
        \vspace{-16pt}
\end{figure}

\section{Model}
\label{sec:Model}
The Faraday effect is usually considered in the collinear geometry where light travels along the magnetic field force lines \cite{Zvedin}. In an initially isotropic optical medium, a simple relation for the angle of the plane of polarization rotation is easily obtained:
\begin{align}
    \Theta_F &= \nu B_{\parallel} L,
    \label{Faraday_rotation_angle_expression}
\end{align}
where $\Theta_F$ is the polarization plane rotation angle, $\nu$ is the Verdet constant, $B_{\parallel}$ is the magnetic field vector $\mathbf{B}$ projection onto the light propagation direction and $L$ is the optical path length. Such interaction geometry has been studied in application to WGMRs in \cite{80Tl}. It has been shown there that a polarization transmission spectrum of a system comprising a WGMR and a coupling element can be easily obtained using a recurrent formula 
\begin{align}
    \boldsymbol{E}_{j+1}=a\hat{M_r} \boldsymbol{E}_j, \label{recurrent}
\end{align}

where $\boldsymbol{E_j}$ is a Jones vector representing the $j-$th roundtrip polarization state, $a=\exp \left( -2n\pi ^{2}R/\lambda Q \right)$ is the extinction coefficient expressed in terms of the quality factor $Q$, $n$ is the refractive index of the medium, $\lambda$ is the light wavelength and $\hat{M}_r$ is the full roundtrip Jones matrix. 
The output Jones vector $\boldsymbol{E}_{out}$ at a given wavelength can be calculated by summarizing the inputs of the waves that have passed through the cavity once, twice, thrice, and so on:
\begin{equation}
\begin{split}
\boldsymbol{E}_{out} = t\boldsymbol{E}_0 &  +\kappa(-\kappa^*) a \hat{M_r}\boldsymbol{E}_0  +\kappa(-\kappa^*) t^* a^2 \hat{M_r}^2\boldsymbol{E}_0  +\dots \\
&=t\boldsymbol{E}_0-\kappa\kappa^* \frac{1}{t^*}\biggl(\sum_{m=1}^{\infty}\bigl(t^*a \hat{M_r}\bigr)^{m}\biggr) \boldsymbol{E}_0,
\label{equation:Esum}
\end{split}
\end{equation}
where $\boldsymbol{E}_0$ is the input light polarization state, $t, t^*, \kappa, \kappa^*$ are the coupling coefficients between the WGMR and the coupling element. Calculating $\boldsymbol{E}_{out}$ with \eqref{equation:Esum} for a range of wavelengths, one can construct the transmission spectrum of a WGMR. The eigenfrequency deviation amplitude is then easily deduced as a function of the applied magnetic field.

The full roundtrip Jones matrix for the collinear configuration can be found using the permittivity tensor \cite{Zvedin}:
\begin{align}
        \hat{\varepsilon}/\varepsilon_0 &= 
        \begin{pmatrix}
            \varepsilon_x & 0 & i\gamma \\
            0  & \varepsilon_y & 0 \\
            -i\gamma & 0 & \varepsilon_z
        \end{pmatrix}. 
        \label{vareps}
\end{align}
Here the coordinate system is chosen so that the magnetic field vector $\boldsymbol{B}$ and the wavevector of light $\boldsymbol{k}$ are co-aligned, the non-diagonal imaginary terms  $\gamma = n \nu \lambda B_{\parallel}/ \pi $, where $n$ is the refractive index, are the components of the gyration vector, representing the Faraday effect \cite{Zvedin}. The diagonal terms are not equal, which accounts for the fact that the eigenfrequencies of the TM (in our model - $x$-polarized) and TE ($z-$polarized) whispering-gallery-modes of the same order are different (Eq. 1 in Ref. \cite{Gorodetsky2}). This difference between the TE and TM modes' eigenfrequencies is calculated, for example, in Ref. \cite{Osada}:
\begin{align}
     \Omega _{TM}-\Omega _{TE}=\frac{c}{Rn^2}\sqrt{n^{2}-1},
     \label{osadas}
\end{align} 
where $c$ is the speed of light, $R$ -- cavity radius,  $\Omega_{TM, \; TE}=2\pi c n_{TM, \; TE}/\lambda$. This difference may be viewed as being caused by the so-called geometric birefringence that can be represented by the permittivity variation $\delta\varepsilon$:
\begin{align}
     \varepsilon_x =& \bar{\varepsilon} + \delta\varepsilon, & 
     \varepsilon_y =& \bar{\varepsilon},  & 
     \varepsilon_z =& \bar{\varepsilon} - \delta\varepsilon,  & 
    n_z-n_x =& \delta n,
     \label{vareps_xyz}
\end{align} 
where $\bar{\varepsilon}$ is simply $n^2$. Henceforth:
\begin{align}
     n_{TM} &=\sqrt{n^2+\delta\varepsilon}, & 
     n_{TE} &=\sqrt{n^2-\delta\varepsilon}.
     \label{n_vareps}
\end{align} 
Combining \eqref{osadas} with \eqref{n_vareps}, one can obtain
\begin{align}
     \delta\varepsilon & \cong \frac{\lambda}{2\pi R}\frac{\sqrt{n^2-1}}{n}.
\end{align}
As shown in Ref. ~\cite{Tabor}, the Jones matrix, for a medium with permittivity tensor $\hat{\varepsilon}$ takes the following form:
\begin{align}
    \hat{M}=
        \begin{pmatrix}
              \cos{\phi}-i \cos{\chi}\sin{\phi} & -\sin{\chi}\sin{\phi}  \\
              \sin{\chi}\sin{\phi} & \cos{\phi}+i \cos{\chi}\sin{\phi} \\
        \end{pmatrix}
        \exp{(-i\psi)}, \label{MJones}
\end{align}
where
\begin{align}
    \psi &=\bar{k} L, & \phi &= \delta k  L,\\
    \delta k &=\frac{k_{TE}-k_{TM}}{2}, & \bar{k} &= \frac{k_{TE}+k_{TM}}{2},\\
    \cos{\chi} &= \frac{\delta\varepsilon}{\sqrt{\delta\varepsilon^2+\gamma^2}}, &
    \sin{\chi} &= \frac{\gamma}{\sqrt{\delta\varepsilon^2+\gamma^2}},
\end{align}
\begin{align}
    k_{TM,TE}^2 &= \bigg(\frac{2 \pi}{\lambda}\bigg)^2 \mu\bigg[\bar{\varepsilon}\pm\sqrt{(\delta \varepsilon)^2+\gamma^2}\bigg]. \label{kTETM}
\end{align}
Here, $L$ is the optical path length, $\mu=1$ due to TGG being paramagnetic. Jones matrix \eqref{MJones} can be applied in the collinear configuration by simply putting $L=2\pi R$, since the angle between the magnetic field vector $\boldsymbol{B}$ and the wavevector $\boldsymbol{k}$ equals zero throughout the whole light path, meaning that ${B}_{\parallel}$ remains constant.

The collinear geometry has been implemented in \cite{Thermally_tunable_WGM}
by passing an electric wire through the center (the symmetry axis) of the optical cavity, as shown in Fig. \ref{fig:field_orientation}(a). The WGM cavity has been formed through melting of a portion of Faraday-rotator material upon a metallic wire. However, our particular goal was to study the possibility of creating a WGMR-based magnetic field sensor, which is why we have turned to the transverse magneto-optic interaction geometry that corresponds to a magnetic field source placed outside the resonator, as shown in Fig. \ref{fig:field_orientation}(b). On the one hand, such configuration is more robust, allows for a stronger magnetic field, and is far easier to implement. On the other hand, the model, described by Eqs. \ref{vareps} to \ref{kTETM}, developed in Ref.~\cite{80Tl}, can not be readily applied in this case since the $B_{\parallel}$ component of the magnetic field $B_{\parallel}$ varies along the light's path. Assuming that the magnetic field is uniform, we can put ${B_{\parallel}}={B}\cos\theta$ for the transverse  configuration (Fig. \ref{fig:field_orientation}(b)). Now we can divide the total circumference of the resonator into $N$ small sections (the upper limit for $N$ is actually the azimuthal number of a WGM, which is in the order of $\sim 10^5$), assuming the angle $\theta =\theta_j$ to be constant within each section. The full roundtrip Jones matrix $\hat{M_r}$ of a WGMR can then be obtained numerically through multiplication of the individual sections' Jones matrices $\hat{M}\big(\theta_j\big)$:
\begin{align}
    \hat{M_r} &= \lim_{N\rightarrow\infty}\prod_{j=1}^{N} \hat{M}\big(\theta_j\big) = \lim_{N\rightarrow\infty}\prod_{j=1}^{N} \hat{M} \bigg(\frac{2\pi j}{N}\bigg). \label{MRound}
\end{align}
This matrix can now be used to calculate the WGMR transmission spectrum using Eq. (\ref{equation:Esum}). It is worth noting that the proposed approach is suitable for arbitrary configurations (e.g. inhomogeneous magnetic field distributions).
The corresponding calculation has been performed using the following set of parameters:  $R=\SI{2.85}{\milli\metre}$, $\nu=\SI[per-mode=symbol]{-111}{\radian\per\tesla\per\metre}$ for $\lambda=\SI{670}{\nano\metre}$ wavelength, $n=1.95$.
The eigenfrequency deviation amplitude calculation results for both configurations are presented in Fig. \ref{fig:Shift}, along with the experimental data for the transverse configuration. The wavevectors' expressions $k_{TM,TE}$ given by \eqref{kTETM} can be expanded in a Taylor series, from which the frequency deviation quadratic dependence on the magnetic field magnitude ($\gamma \sim B_{\parallel}$) is clear:
\begin{align}
    k_{TM,TE} &= \frac{2 \pi}{\lambda} \bigg( \sqrt{\bar{\varepsilon} \pm \delta \varepsilon}\pm 
    \frac{ \gamma ^2}{4 \delta \varepsilon \sqrt{\bar{\varepsilon} \pm \delta \varepsilon}} + O(\gamma^4) \bigg).
\end{align}
The resulting tuning rates are $\SI[per-mode=symbol]{1.25e-6}{\mega\hertz\per\gauss\squared}$ for the transverse configuration and $\SI[per-mode=symbol]{10.43e-6}{\mega\hertz\per\gauss\squared}$ for the collinear configuration. The utilized scheme (\ref{MRound}) was found to converge as fast as $1/N^2$.

\section{Experimental results}
\label{sec:experiment}
\subsection{Frequency deviation amplitude measurements}

The $Q$-factor of the $\SI{2.85}{\milli\metre}$-diameter WGMR fabricated of TGG (see Section 1 of Supplement for details on fabrication process) has been estimated to be $(1.45\pm 0.17) \times 10^8$ for the TE-modes using the technique developed in Ref. \cite{Matsko} (see Section 2 of Supplement), which can be considered as a record among TGG materials bearing high Verdet constant.

\begin{figure}[bth]
\centering
\includegraphics[width=1\linewidth]{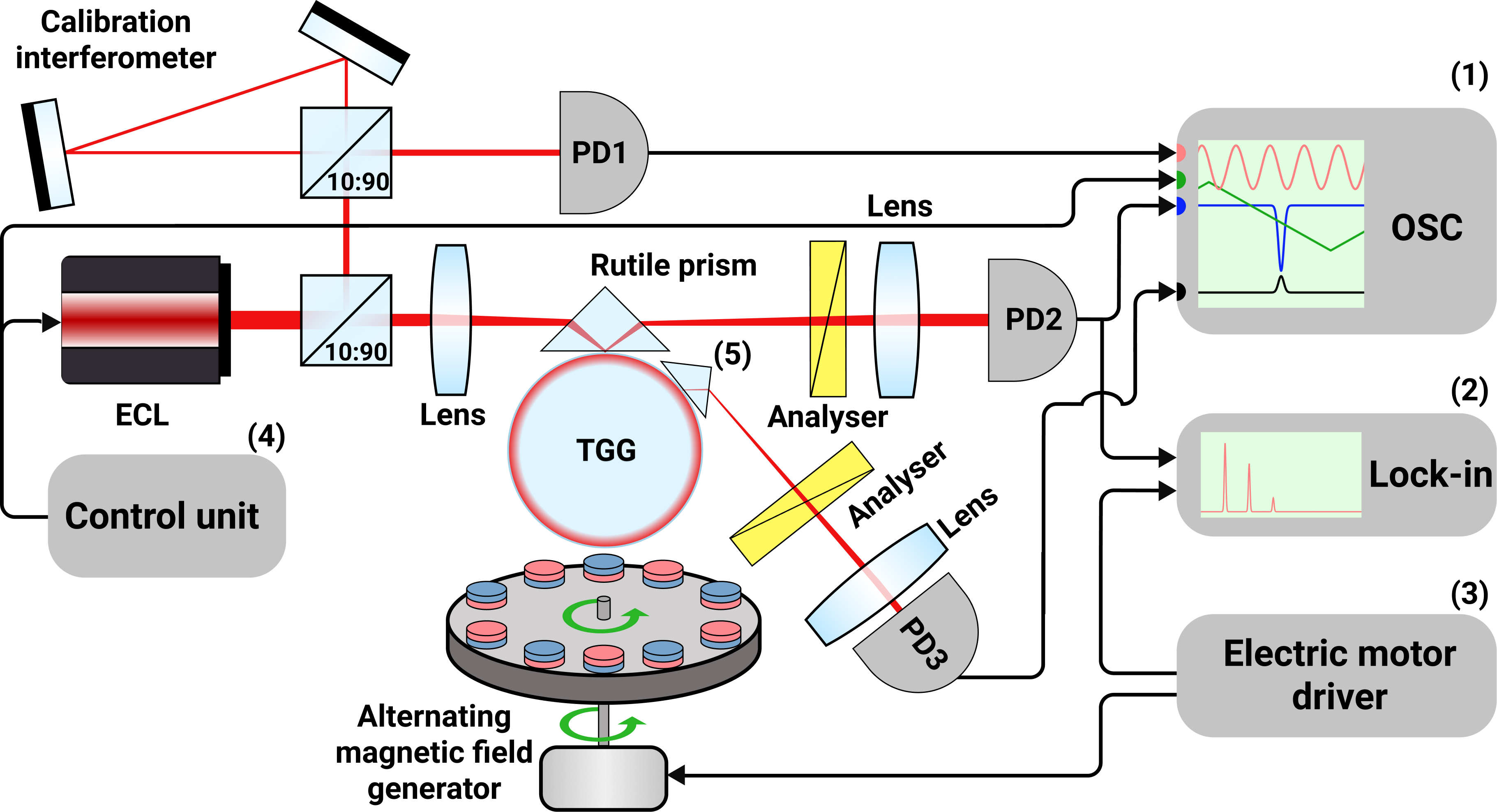}
\caption{Experimental setup: (1) -- oscilloscope; (2) -- lock-in amplifier; (3) -- stepper motor driver; (4) -- external cavity laser control unit; (5) -- diamond prism; PD 1-3 -- photodetectors; TGG -- microresonator; Analyser -- Glann prism polarization analyser.
\label{setup_main}}
\vspace{-10pt}
\end{figure}

A schematic of the experimental setup is presented in Fig. \ref{setup_main}. The light from a tunable ($667 \leq \lambda \leq \SI{671}{\nano\metre}$) $\SI{35}{\milli\watt}$ external-cavity laser (ECL) is focused onto an inner surface of a rutile ($TiO_2$) prism via a lens. The refractive index for the extraordinary wave (TE mode) in the prism is $n_{c} = 2.95$ (the refractive index of TGG is $1.95$). A small portion of the light beam intensity has been coupled into a $\SI{89}{\mega\hertz}$ FSR free-space interferometer to provide control over the laser wavelength sweep (PD1). 
Lastly, the light leaving the WGMR through the prism is focused onto the aperture of a photodetector (PD2).

\begin{figure}[bth]
    \centering
    \includegraphics[width=1\linewidth]{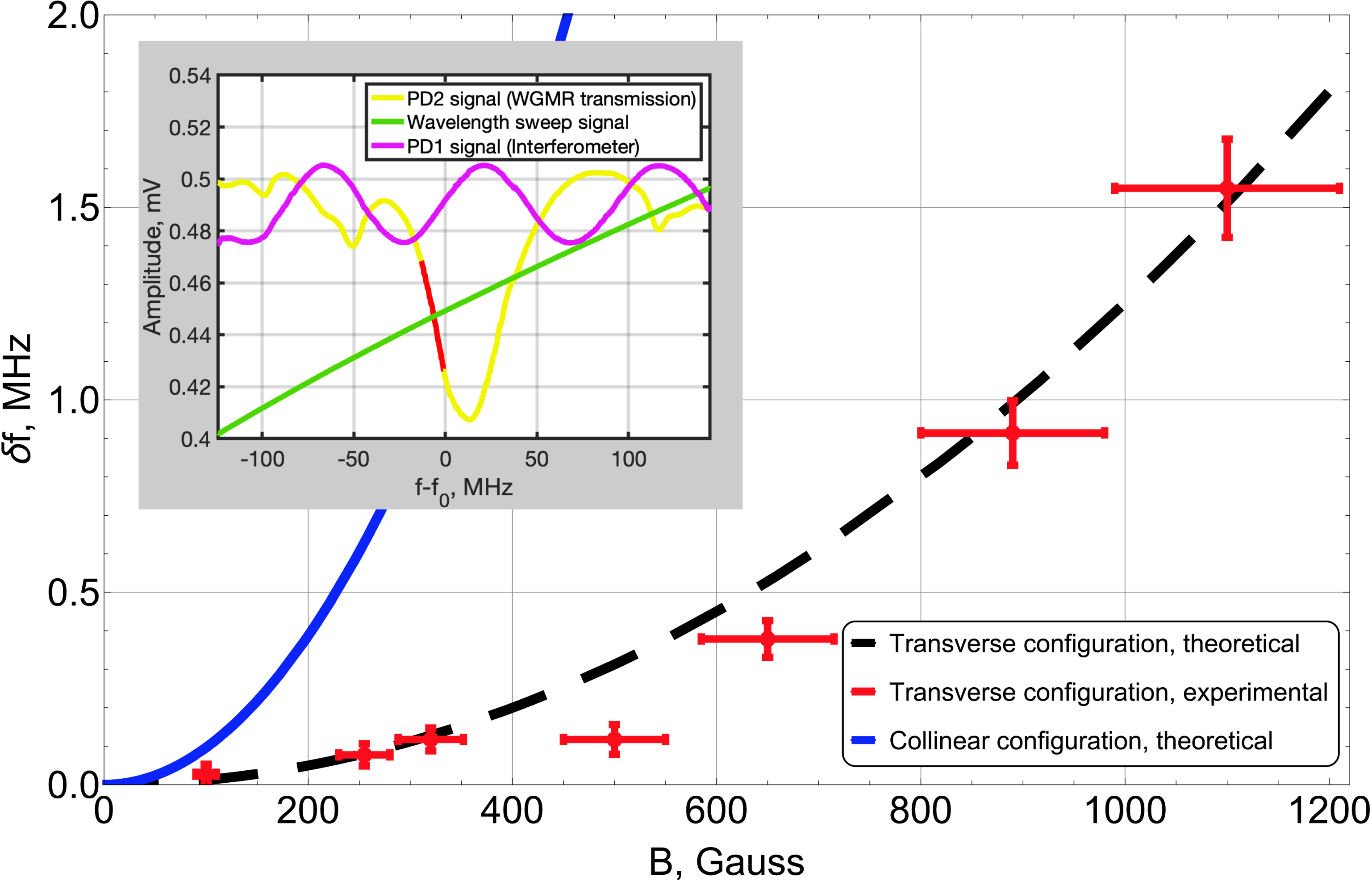}
    \caption{The TGG WGMR eigenmode frequency modulation amplitude on the second harmonic of the applied  magnetic field. The inset shows the WGMR transmission spectrum obtained in the wavelength sweep mode of the laser. The traces are: yellow -- PD2 signal, magenta -- PD1 signal, green -- laser wavelength sweep signal. The markers highlight the slope section used in the experiment.
    \vspace{-16pt}
    \label{fig:Shift}}
\end{figure}

A high-amplitude alternating magnetic field generator has been constructed (Fig. \ref{setup_main}) by placing ten permanent magnets on a rotating disk driven with a low mechanical noise stepper motor, each neighboring magnet facing the opposite pole to the resonator. The resulting alternating magnetic field signal was close to sinusoidal with a negligible harmonics coefficient. The magnetic field magnitude was set by adjusting the distance between the disk and the WGMR and measured with a Hall sensor placed in the proximity of the cavity. Dynamic measurements were performed at magnetic field alternation frequencies ranging from $20$ to $\SI{40}{\hertz}$. 
The laser was tuned to the resonance dip slope (Fig. \ref{fig:Shift}, inset), which allowed to translate its side-to side movement caused by the eigenfrequency deviation into the PD2 signal amplitude variation. This signal was then fed to a lock-in amplifier (Stanford Research Systems SR560) along with the magnetic field generator driving signal, which allowed to drastically increase the setup sensitivity (see Section 3 of Supplement). 

The resulting frequency deviation amplitude measured on the second harmonic of the driving signal is plotted against the magnetic field magnitude in Fig. \ref{fig:Shift}. The experimental result (dots) shows good correspondence with the theoretical predictions (black dashed line). The x-axis error comes from the Hall sensor positioning uncertainty, while the y-axis precision is limited by the slow thermal drift of the system (see Section 3 of Supplement). The frequency deviation amplitude was found to be of $\SI{1.6}{\mega\hertz}$ at the maximum achievable in this setup field strength of $\SI{1100}{G}$.

\subsection{Polarization rotation measurements}
Let us consider a hypothetical case of a WGMR that possesses no geometrical birefringence (the TE and TM modes frequencies are equal). In this case, the resulting plane of polarization rotation angle $\Theta_r$, accumulated throughout a full light roundtrip, is simply a line integral of the magnetic field vector $\boldsymbol{B}$ projection onto the light propagation direction.
\begin{align}
    \Theta_r &= \nu R \int_{0}^{2\pi}  \frac{(\boldsymbol{B} \cdot \boldsymbol{k})}{|\boldsymbol{k}|} d\theta.
\end{align}
This way, since the light path within the optical cavity is a closed-loop, in a non-birefringent WGMR the polarization rotation angle caused by the Faraday effect, will be precisely zero due to Ampere's circuital law in the transverse geometry.

Even in the real case where the Faraday rotation is hindered by the geometrical birefringence \cite{80Tl}, one can reasonably expect that the full roundtrip polarization plane rotation angle will also be zero, or, at least, close to zero. To test this hypothesis, polarization state measurements have been performed. An analyzer has been placed between the collimating lens and PD2 to probe the throughput light polarization state. The magnetic field was created by a single permanent magnet in this case. As expected, no polarization plane rotation (TE and TM modes mixing) has been observed. The accuracy of these measurements has been estimated to be $\sim \ang{0.01}$, limited by the minimum resolvable PD2 signal variation.

\begin{figure}[htb] \centering \includegraphics[width=1\linewidth]{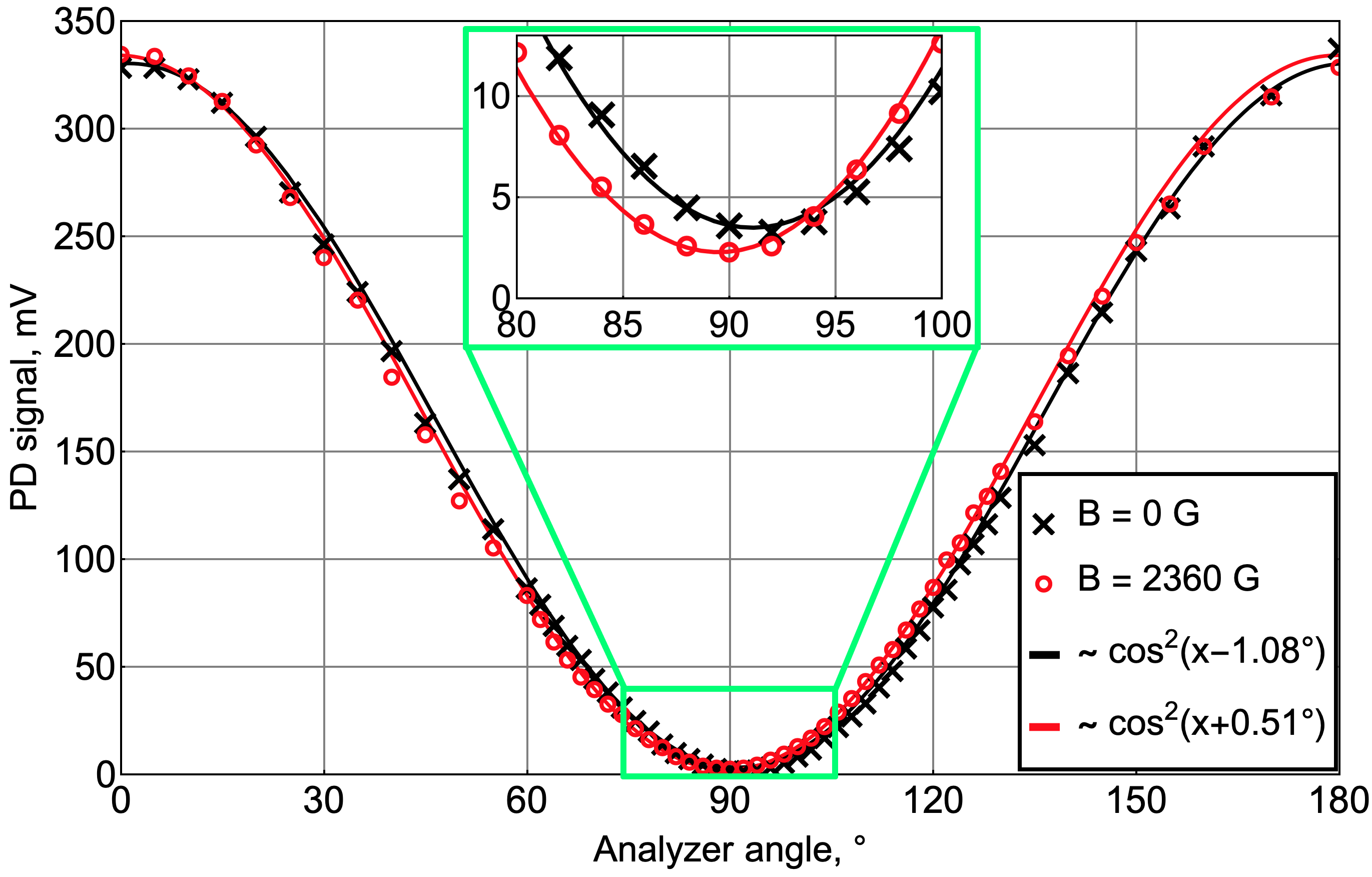} \caption{
		Measured intensity of the light at the resonant wavelength, out-coupled by the additional diamond prism located $\ang{89}$ away from the coupling prism, vs. polarizer angle. A polarization plane rotation of $\ang{1.59}$ is observed.  }
    \label{fig:polarization_Rotation}
    \vspace{-16pt}
\end{figure}

In order to further support our rotation cancellation hypothesis it was decided to probe the polarization state at some intermediate point by using an additional coupling element -- a diamond prism, which allowed us to assess the out-coupled light polarization state, as shown in Fig. \ref{setup_main}. The diamond prism output light intensity (PD3) is plotted against the analyzer rotation angle in Fig. \ref{fig:polarization_Rotation}. 
It follows the well-known Malus law pattern for both cases (with and without magnetic field application). 
The amount of polarization plane rotation acquired along the light path section ($\ang{89}$) between the rutile and the diamond prisms is the difference between the x-axis positions of the $\cos^2$ curves and it equals $\ang{1.59}$ declination from the initial state. A calculation performed via Eq. (\ref{equation:Esum}) gives us $\ang{1.53}$ declination, in good correspondence with the experimental results. We believe that this validates our previous result -- the absence of polarization plane rotation for a full roundtrip. 

\section{Discussion}

We have demonstrated to the best of our knowledge the record quality factor ($Q =1.45\times 10^8$) among materials with a high Verdet constant.

We have observed an eigenfrequency modulation, induced by a harmonic low-frequency transverse homogeneous magnetic field. The eigenfrequency deviation amplitude turned out to be proportional to the magnetic field magnitude squared. 
The resulting frequency deviation amplitude is independent of the $Q$-factor. Nevertheless, a high $Q$-factor allows us to increase the measurement accuracy. A simple model based on the Jones matrices product calculation has been developed to describe this behavior. 
The calculation results are in excellent agreement with the experimental data. It is important that the developed model is applicable for arbitrary magnetic field distributions.

As a weak magnetic field sensor, TGG WGMR is, obviously, inferior to many other devices, offering sensitivity below $\SI{1}{\femto \tesla}$ \cite{SERF} or superior field strength to frequency deviation amplitude conversion coefficient at low frequencies \cite{yu2016optomechanical}. 
Nevertheless, TGG WGMR has its advantages, including higher frequency field detection capability, limited only by the optical mode relaxation time ($\SI{3}{\mega\hertz}$ in our case), and absence of saturation at any reachable magnetic field strengths. The possible utilization of the interaction between alternatively polarized (TE-like and TM-like) mode families, via a high-frequency alternating magnetic field in WGM resonators, manufactured from Faraday-rotating materials deserves further study.

\section*{Funding}
This work was supported by the Russian Science Foundation (project 20-12-00344).

\section*{Acknowledgments}
Verdet constant for the used TGG samples was measured at the VNIIOFI center of the shared use (VNIIOFI CSU). The authors thank Artem E. Shitikov for useful discussion. 

%\section*{Disclosures}
%The authors declare no conflicts of interest.

\section*{Supplemental document}
See \href{https://osapublishing.figshare.com/s/d8de0a651a1f2f497cd5}{Supplement 1} for supporting content.

\bibliography{main}

\providecommand{\noopsort}[1]{}\providecommand{\singleletter}[1]{#1}%
\begin{thebibliography}{26}
\expandafter\ifx\csname natexlab\endcsname\relax\def\natexlab#1{#1}\fi
\expandafter\ifx\csname bibnamefont\endcsname\relax
  \def\bibnamefont#1{#1}\fi
\expandafter\ifx\csname bibfnamefont\endcsname\relax
  \def\bibfnamefont#1{#1}\fi
\expandafter\ifx\csname citenamefont\endcsname\relax
  \def\citenamefont#1{#1}\fi
\expandafter\ifx\csname url\endcsname\relax
  \def\url#1{\texttt{#1}}\fi
\expandafter\ifx\csname urlprefix\endcsname\relax\def\urlprefix{URL }\fi
\providecommand{\bibinfo}[2]{#2}
\providecommand{\eprint}[2][]{\url{#2}}

\bibitem[{\citenamefont{Braginsky et~al.}(1989)\citenamefont{Braginsky,
  Gorodetsky, and Ilchenko}}]{BGII}
\bibinfo{author}{\bibfnamefont{V.}~\bibnamefont{Braginsky}},
  \bibinfo{author}{\bibfnamefont{M.}~\bibnamefont{Gorodetsky}},
  \bibnamefont{and} \bibinfo{author}{\bibfnamefont{V.}~\bibnamefont{Ilchenko}},
  \bibinfo{journal}{Physics Letters A.} \textbf{\bibinfo{volume}{137(7-8)}},
  \bibinfo{pages}{393} (\bibinfo{year}{1989}).

\bibitem[{\citenamefont{Grudinin et~al.}(2006)\citenamefont{Grudinin, Matsko,
  Savchenkov, Strekalov, Ilchenko, and
  Maleki}}]{Ultra_high_Q_crystalline_microcavities}
\bibinfo{author}{\bibfnamefont{I.}~\bibnamefont{Grudinin}},
  \bibinfo{author}{\bibfnamefont{A.}~\bibnamefont{Matsko}},
  \bibinfo{author}{\bibfnamefont{A.}~\bibnamefont{Savchenkov}},
  \bibinfo{author}{\bibfnamefont{D.}~\bibnamefont{Strekalov}},
  \bibinfo{author}{\bibfnamefont{V.}~\bibnamefont{Ilchenko}}, \bibnamefont{and}
  \bibinfo{author}{\bibfnamefont{.~L.} \bibnamefont{Maleki}},
  \bibinfo{journal}{Optics Communications} pp. \bibinfo{pages}{33--38}
  (\bibinfo{year}{2006}).

\bibitem[{\citenamefont{Ward and
  Benson}(2011)}]{WGM_microresonators_sensing_lasing}
\bibinfo{author}{\bibfnamefont{J.}~\bibnamefont{Ward}} \bibnamefont{and}
  \bibinfo{author}{\bibfnamefont{O.}~\bibnamefont{Benson}},
  \bibinfo{journal}{Laser \& Photonics Reviews} pp. \bibinfo{pages}{553--570}
  (\bibinfo{year}{2011}).

\bibitem[{\citenamefont{Strekalov et~al.}(2016)\citenamefont{Strekalov,
  Marquardt, Matsko, Schwefel, and Leuchs}}]{Nonlinear_and_quantum_optics}
\bibinfo{author}{\bibfnamefont{D.}~\bibnamefont{Strekalov}},
  \bibinfo{author}{\bibfnamefont{C.}~\bibnamefont{Marquardt}},
  \bibinfo{author}{\bibfnamefont{A.}~\bibnamefont{Matsko}},
  \bibinfo{author}{\bibfnamefont{H.}~\bibnamefont{Schwefel}}, \bibnamefont{and}
  \bibinfo{author}{\bibfnamefont{G.}~\bibnamefont{Leuchs}},
  \bibinfo{journal}{Journal of Optics}  (\bibinfo{year}{2016}).

\bibitem[{\citenamefont{Matsko and Ilchenko}(2006)}]{Optical_resonators_part1}
\bibinfo{author}{\bibfnamefont{A.}~\bibnamefont{Matsko}} \bibnamefont{and}
  \bibinfo{author}{\bibfnamefont{V.}~\bibnamefont{Ilchenko}},
  \bibinfo{journal}{IEEE Journal of Selected Topics in Quantum Electronics}
  \textbf{\bibinfo{volume}{12}}, \bibinfo{pages}{3} (\bibinfo{year}{2006}).

\bibitem[{\citenamefont{Ilchenko and Matsko}(2006)}]{Optical_resonators_part2}
\bibinfo{author}{\bibfnamefont{V.}~\bibnamefont{Ilchenko}} \bibnamefont{and}
  \bibinfo{author}{\bibfnamefont{A.}~\bibnamefont{Matsko}},
  \bibinfo{journal}{IEEE Journal of Selected Topics in Quantum Electronics}
  \textbf{\bibinfo{volume}{12}}, \bibinfo{pages}{15} (\bibinfo{year}{2006}).

\bibitem[{\citenamefont{Savchenkov et~al.}(2004)\citenamefont{Savchenkov,
  Ilchenko, Matsko, and Maleki}}]{Kilohertz_optical_resonances}
\bibinfo{author}{\bibfnamefont{A.}~\bibnamefont{Savchenkov}},
  \bibinfo{author}{\bibfnamefont{V.}~\bibnamefont{Ilchenko}},
  \bibinfo{author}{\bibfnamefont{A.}~\bibnamefont{Matsko}}, \bibnamefont{and}
  \bibinfo{author}{\bibfnamefont{L.}~\bibnamefont{Maleki}},
  \bibinfo{journal}{Physical Review A}  (\bibinfo{year}{2004}).

\bibitem[{\citenamefont{Bogatyrev et~al.}(2002)\citenamefont{Bogatyrev,
  Vovchenko, Krasyuk, Oboev, Semenov, and Sychugov}}]{Acoustic_sensor}
\bibinfo{author}{\bibfnamefont{V.}~\bibnamefont{Bogatyrev}},
  \bibinfo{author}{\bibfnamefont{V.}~\bibnamefont{Vovchenko}},
  \bibinfo{author}{\bibfnamefont{I.}~\bibnamefont{Krasyuk}},
  \bibinfo{author}{\bibfnamefont{V.}~\bibnamefont{Oboev}},
  \bibinfo{author}{\bibfnamefont{A.}~\bibnamefont{Semenov}}, \bibnamefont{and}
  \bibinfo{author}{\bibfnamefont{V.}~\bibnamefont{Sychugov},
  \bibfnamefont{VA~Torchigin}}, \bibinfo{journal}{Quantum Electronics}
  \textbf{\bibinfo{volume}{32}}, \bibinfo{pages}{471} (\bibinfo{year}{2002}).

\bibitem[{\citenamefont{Guan et~al.}(2006)\citenamefont{Guan, Arnold, and
  Otugen}}]{Temp_sensor}
\bibinfo{author}{\bibfnamefont{G.}~\bibnamefont{Guan}},
  \bibinfo{author}{\bibfnamefont{S.}~\bibnamefont{Arnold}}, \bibnamefont{and}
  \bibinfo{author}{\bibfnamefont{M.}~\bibnamefont{Otugen}},
  \bibinfo{journal}{AIAA Journal} \textbf{\bibinfo{volume}{44}},
  \bibinfo{pages}{2385} (\bibinfo{year}{2006}).

\bibitem[{\citenamefont{Ma et~al.}(2010)\citenamefont{Ma, Huang, and
  Guo}}]{Humidity_sensor}
\bibinfo{author}{\bibfnamefont{Q.}~\bibnamefont{Ma}},
  \bibinfo{author}{\bibfnamefont{L.}~\bibnamefont{Huang}}, \bibnamefont{and}
  \bibinfo{author}{\bibfnamefont{T.}~\bibnamefont{Guo},
  \bibfnamefont{Z~Rossmann}}, \bibinfo{journal}{Measurement Science and
  Technology} \textbf{\bibinfo{volume}{21}}, \bibinfo{pages}{115206}
  (\bibinfo{year}{2010}).

\bibitem[{\citenamefont{Huston and Eversole}(1993)}]{fiberstrain}
\bibinfo{author}{\bibfnamefont{A.}~\bibnamefont{Huston}} \bibnamefont{and}
  \bibinfo{author}{\bibfnamefont{J.}~\bibnamefont{Eversole}},
  \bibinfo{journal}{Optics Letters} \textbf{\bibinfo{volume}{18}},
  \bibinfo{pages}{1104} (\bibinfo{year}{1993}).

\bibitem[{\citenamefont{Wang et~al.}(2015)\citenamefont{Wang, Bo, Wan, Li, Gao,
  Li, Zhang, and J.}}]{electroopticalmodulation}
\bibinfo{author}{\bibfnamefont{J.}~\bibnamefont{Wang}},
  \bibinfo{author}{\bibfnamefont{F.}~\bibnamefont{Bo}},
  \bibinfo{author}{\bibfnamefont{S.}~\bibnamefont{Wan}},
  \bibinfo{author}{\bibfnamefont{W.}~\bibnamefont{Li}},
  \bibinfo{author}{\bibfnamefont{F.}~\bibnamefont{Gao}},
  \bibinfo{author}{\bibfnamefont{J.}~\bibnamefont{Li}},
  \bibinfo{author}{\bibfnamefont{G.}~\bibnamefont{Zhang}}, \bibnamefont{and}
  \bibinfo{author}{\bibfnamefont{X.}~\bibnamefont{J.}},
  \bibinfo{journal}{Optics Express} \textbf{\bibinfo{volume}{23}},
  \bibinfo{pages}{23072} (\bibinfo{year}{2015}).

\bibitem[{\citenamefont{Yu et~al.}(2016)\citenamefont{Yu, Janousek, Sheridan,
  McAuslan, Rubinsztein-Dunlop, Lam, Zhang, and Bowen}}]{yu2016optomechanical}
\bibinfo{author}{\bibfnamefont{C.}~\bibnamefont{Yu}},
  \bibinfo{author}{\bibfnamefont{J.}~\bibnamefont{Janousek}},
  \bibinfo{author}{\bibfnamefont{E.}~\bibnamefont{Sheridan}},
  \bibinfo{author}{\bibfnamefont{D.~L.} \bibnamefont{McAuslan}},
  \bibinfo{author}{\bibfnamefont{H.}~\bibnamefont{Rubinsztein-Dunlop}},
  \bibinfo{author}{\bibfnamefont{P.~K.} \bibnamefont{Lam}},
  \bibinfo{author}{\bibfnamefont{Y.}~\bibnamefont{Zhang}}, \bibnamefont{and}
  \bibinfo{author}{\bibfnamefont{W.~P.} \bibnamefont{Bowen}},
  \bibinfo{journal}{Physical Review Applied} \textbf{\bibinfo{volume}{5}},
  \bibinfo{pages}{044007} (\bibinfo{year}{2016}).

\bibitem[{\citenamefont{Villaverde et~al.}(1978)\citenamefont{Villaverde,
  Donatti, and Bozinis}}]{TGG_pulsed}
\bibinfo{author}{\bibfnamefont{A.}~\bibnamefont{Villaverde}},
  \bibinfo{author}{\bibfnamefont{D.}~\bibnamefont{Donatti}}, \bibnamefont{and}
  \bibinfo{author}{\bibfnamefont{D.}~\bibnamefont{Bozinis}},
  \bibinfo{journal}{Journal of Physics C} \textbf{\bibinfo{volume}{11}},
  \bibinfo{pages}{L495} (\bibinfo{year}{1978}).

\bibitem[{\citenamefont{Slezák et~al.}(2016)\citenamefont{Slezák, Yasuhara,
  Lucianetti, and Mocek}}]{TGG_temp_wave}
\bibinfo{author}{\bibfnamefont{O.}~\bibnamefont{Slezák}},
  \bibinfo{author}{\bibfnamefont{R.}~\bibnamefont{Yasuhara}},
  \bibinfo{author}{\bibfnamefont{A.}~\bibnamefont{Lucianetti}},
  \bibnamefont{and} \bibinfo{author}{\bibfnamefont{T.}~\bibnamefont{Mocek}},
  \bibinfo{journal}{Optical Materials Express} \textbf{\bibinfo{volume}{6}},
  \bibinfo{pages}{3683} (\bibinfo{year}{2016}).

\bibitem[{\citenamefont{Cruz et~al.}(1996)\citenamefont{Cruz, Andres, and
  Hernandez}}]{SilicaFibreFaraday}
\bibinfo{author}{\bibfnamefont{J.}~\bibnamefont{Cruz}},
  \bibinfo{author}{\bibfnamefont{M.}~\bibnamefont{Andres}}, \bibnamefont{and}
  \bibinfo{author}{\bibfnamefont{M.}~\bibnamefont{Hernandez}},
  \bibinfo{journal}{Applied Optics} \textbf{\bibinfo{volume}{35}},
  \bibinfo{pages}{922} (\bibinfo{year}{1996}).

\bibitem[{\citenamefont{V{\'\i}llora et~al.}(2011)\citenamefont{V{\'\i}llora,
  Molina, Nakamura, Shimamura, Hatanaka, Funaki, and Naoe}}]{TGG_TSLAG_FOM}
\bibinfo{author}{\bibfnamefont{E.~G.} \bibnamefont{V{\'\i}llora}},
  \bibinfo{author}{\bibfnamefont{P.}~\bibnamefont{Molina}},
  \bibinfo{author}{\bibfnamefont{M.}~\bibnamefont{Nakamura}},
  \bibinfo{author}{\bibfnamefont{K.}~\bibnamefont{Shimamura}},
  \bibinfo{author}{\bibfnamefont{T.}~\bibnamefont{Hatanaka}},
  \bibinfo{author}{\bibfnamefont{A.}~\bibnamefont{Funaki}}, \bibnamefont{and}
  \bibinfo{author}{\bibfnamefont{K.}~\bibnamefont{Naoe}},
  \bibinfo{journal}{Applied Physics Letters} \textbf{\bibinfo{volume}{99}},
  \bibinfo{pages}{011111} (\bibinfo{year}{2011}).

\bibitem[{\citenamefont{Bai et~al.}(2003)\citenamefont{Bai, Lu, and
  Lin}}]{YIG_FOM}
\bibinfo{author}{\bibfnamefont{J.~G.} \bibnamefont{Bai}},
  \bibinfo{author}{\bibfnamefont{G.-Q.} \bibnamefont{Lu}}, \bibnamefont{and}
  \bibinfo{author}{\bibfnamefont{T.}~\bibnamefont{Lin}},
  \bibinfo{journal}{Sensors and Actuators A: Physical}
  \textbf{\bibinfo{volume}{109}}, \bibinfo{pages}{9} (\bibinfo{year}{2003}).

\bibitem[{\citenamefont{Zvezdin and Kotov}(1997)}]{Zvedin}
\bibinfo{author}{\bibfnamefont{A.~K.} \bibnamefont{Zvezdin}} \bibnamefont{and}
  \bibinfo{author}{\bibfnamefont{V.~A.} \bibnamefont{Kotov}},
  \emph{\bibinfo{title}{{Modern magnetooptics and magnetooptical materials}}}
  (\bibinfo{publisher}{CRC Press}, \bibinfo{year}{1997}).

\bibitem[{\citenamefont{Lan and Hossein-Zadeh}(2011)}]{80Tl}
\bibinfo{author}{\bibfnamefont{S.}~\bibnamefont{Lan}} \bibnamefont{and}
  \bibinfo{author}{\bibfnamefont{M.}~\bibnamefont{Hossein-Zadeh}},
  \bibinfo{journal}{IEEE Photonics Journal} \textbf{\bibinfo{volume}{3}},
  \bibinfo{pages}{872} (\bibinfo{year}{2011}).

\bibitem[{\citenamefont{Gorodetsky and Fomin}(2006)}]{Gorodetsky2}
\bibinfo{author}{\bibfnamefont{M.~L.} \bibnamefont{Gorodetsky}}
  \bibnamefont{and} \bibinfo{author}{\bibfnamefont{A.~E.} \bibnamefont{Fomin}},
  \bibinfo{journal}{IEEE Journal of Selected Topics in Quantum Electronics}
  \textbf{\bibinfo{volume}{12}}, \bibinfo{pages}{33} (\bibinfo{year}{2006}).

\bibitem[{\citenamefont{Osada et~al.}(2016)\citenamefont{Osada, Hisatomi,
  Noguchi, Tabuchi, Yamazaki, Usami, Sadgrove, Yalla, Nomura, and
  Nakamura}}]{Osada}
\bibinfo{author}{\bibfnamefont{A.}~\bibnamefont{Osada}},
  \bibinfo{author}{\bibfnamefont{R.}~\bibnamefont{Hisatomi}},
  \bibinfo{author}{\bibfnamefont{A.}~\bibnamefont{Noguchi}},
  \bibinfo{author}{\bibfnamefont{Y.}~\bibnamefont{Tabuchi}},
  \bibinfo{author}{\bibfnamefont{R.}~\bibnamefont{Yamazaki}},
  \bibinfo{author}{\bibfnamefont{K.}~\bibnamefont{Usami}},
  \bibinfo{author}{\bibfnamefont{M.}~\bibnamefont{Sadgrove}},
  \bibinfo{author}{\bibfnamefont{R.}~\bibnamefont{Yalla}},
  \bibinfo{author}{\bibfnamefont{M.}~\bibnamefont{Nomura}}, \bibnamefont{and}
  \bibinfo{author}{\bibfnamefont{Y.}~\bibnamefont{Nakamura}},
  \bibinfo{journal}{Physical Review Letters} \textbf{\bibinfo{volume}{116}},
  \bibinfo{pages}{223601} (\bibinfo{year}{2016}).

\bibitem[{\citenamefont{Tabor and Chen}(1969)}]{Tabor}
\bibinfo{author}{\bibfnamefont{W.}~\bibnamefont{Tabor}} \bibnamefont{and}
  \bibinfo{author}{\bibfnamefont{F.}~\bibnamefont{Chen}},
  \bibinfo{journal}{Journal of Applied Physics} \textbf{\bibinfo{volume}{40}},
  \bibinfo{pages}{2760} (\bibinfo{year}{1969}).

\bibitem[{\citenamefont{Vincent et~al.}(2020)\citenamefont{Vincent, Jiang,
  Russell, and Vollmer}}]{Thermally_tunable_WGM}
\bibinfo{author}{\bibfnamefont{S.}~\bibnamefont{Vincent}},
  \bibinfo{author}{\bibfnamefont{X.}~\bibnamefont{Jiang}},
  \bibinfo{author}{\bibfnamefont{P.}~\bibnamefont{Russell}}, \bibnamefont{and}
  \bibinfo{author}{\bibfnamefont{F.}~\bibnamefont{Vollmer}},
  \bibinfo{journal}{Applied Physics Letters} \textbf{\bibinfo{volume}{116}},
  \bibinfo{pages}{161110} (\bibinfo{year}{2020}).

\bibitem[{\citenamefont{Savchenkov et~al.}(2018)\citenamefont{Savchenkov,
  Borri, de~Cumis, Matsko, De~Natale, and Maleki}}]{Matsko}
\bibinfo{author}{\bibfnamefont{A.~A.} \bibnamefont{Savchenkov}},
  \bibinfo{author}{\bibfnamefont{S.}~\bibnamefont{Borri}},
  \bibinfo{author}{\bibfnamefont{M.~S.} \bibnamefont{de~Cumis}},
  \bibinfo{author}{\bibfnamefont{A.~B.} \bibnamefont{Matsko}},
  \bibinfo{author}{\bibfnamefont{P.}~\bibnamefont{De~Natale}},
  \bibnamefont{and} \bibinfo{author}{\bibfnamefont{L.}~\bibnamefont{Maleki}},
  \bibinfo{journal}{Applied Physics B} \textbf{\bibinfo{volume}{124}},
  \bibinfo{pages}{171} (\bibinfo{year}{2018}).

\bibitem[{\citenamefont{Dang et~al.}(2010)\citenamefont{Dang, Maloof, and
  Romalis}}]{SERF}
\bibinfo{author}{\bibfnamefont{H.}~\bibnamefont{Dang}},
  \bibinfo{author}{\bibfnamefont{A.~C.} \bibnamefont{Maloof}},
  \bibnamefont{and} \bibinfo{author}{\bibfnamefont{M.~V.}
  \bibnamefont{Romalis}}, \bibinfo{journal}{Applied Physics Letters}
  \textbf{\bibinfo{volume}{97}}, \bibinfo{pages}{151110}
  (\bibinfo{year}{2010}).

\end{thebibliography}

\end{document}